\documentclass[aps,prb,reprint,footinbib]{revtex4-2}
\usepackage{physics,amssymb,nicematrix,bm,setspace,microtype,siunitx,capt-of,mathtools,hyperref}
\hypersetup{hidelinks}

\newlength{\mwidth}

\newcommand{\nvec}{\mathmakebox[\mwidth][c]{\bm{n}}}
\newlength{\idxwidth}
\newcommand{\idx}[1]{\mathmakebox[\idxwidth][c]{\scriptstyle #1}}

\newcommand{\eq}[1]{\text{$#1$}}
\newcommand{\mr}[1]{\mathrm{#1}}

\begin{document}
\title{Phenomenology of reactive and dissipative nutation in antiferromagnets}
\author{Erlend Sylju{\aa}sen}
\affiliation{Center for Quantum Spintronics, Department of Physics, Norwegian University of Science and Technology, NO-7491 Trondheim, Norway}
\author{Hans Gl{\o}ckner Giil}
\affiliation{Center for Quantum Spintronics, Department of Physics, Norwegian University of Science and Technology, NO-7491 Trondheim, Norway}
\author{Arne Brataas}
\affiliation{Center for Quantum Spintronics, Department of Physics, Norwegian University of Science and Technology, NO-7491 Trondheim, Norway}

\begin{abstract}
We formulate equations of motion for uniform magnetization dynamics in multisublattice magnets, with emphasis on a recently proposed inertial damping torque. We include intra- and intersublattice contributions to Gilbert damping, reactive inertia, and inertial damping. For a two-sublattice uniaxial antiferromagnet in an easy-axis static magnetic field, we derive the dynamical magnetic susceptibilities and approximate eigenfrequencies across all field-dependent equilibrium phases. In particular, we show how inertial damping affects the relaxation of the nutational modes and propose an experimental protocol to determine the dynamical coefficients.
\end{abstract}
\maketitle

\section{Introduction}
\vspace{-0.2cm}
The Landau-Lifshitz-Gilbert equation provides the standard description of magnetization precession and damping \cite{landau_1935,gilbert_2004}. This may become insufficient at short timescales, where the magnetization dynamics can show additional high-frequency motion, such as nutation superimposed on the ordinary precession \cite{ciornei_2011, neeraj_2021}. Understanding these high-frequency excitations is important for ultrafast magnetic spectroscopy \cite{mondal_2023} and for emerging probes of magnetic dynamics including quantum sensors \cite{heitzer_2024}. The high-frequency motion originates from higher-order time derivatives beyond the standard first-order dynamics. Previous studies have mainly focused on a mass-like inertial contribution \cite{ciornei_2011, fahnle_2011, bhattacharjee_2012, mondal_2023}. This gives rise to reactive (non-dissipative) nutational dynamics, while the damping of the high-frequency motion has often been treated only through the usual Gilbert term.

More recently, an inertial damping torque was proposed for a ferromagnet in Ref.~\cite{giil_2026}. Its relevance is supported by experiments in Refs.~\cite{neeraj_2021,unikandanunni_2022}, where nutational resonances were found to be more strongly damped than expected from Gilbert damping alone. This points to an additional dissipative contribution to the nutational dynamics, which can be captured by the inertial damping torque~\cite{giil_2026}. The nutational frequency is then primarily set by the reactive inertial term, whereas its linewidth also depends on both Gilbert and inertial damping~\cite{giil_2026}. Including the inertial damping torque is therefore important for describing nutational modes.

Antiferromagnets provide a natural setting for studying inertial effects \cite{mondal_2021,mondal_2021a,mondal_2022,winter_2022,titov_2024,angster_2025}. Their exchange-driven magnetic dynamics can occur at higher characteristic frequencies than in ferromagnets, making inertial corrections to the magnetization dynamics more relevant. An applied static magnetic field can also change the equilibrium magnetic configuration, allowing inertial effects to be studied in different magnetic phases. Furthermore, the presence of multiple magnetic sublattices permits dynamical coupling between them. This can give rise to additional modes whose frequencies and linewidths depend on both intra- and intersublattice dynamical coefficients. Such contributions have previously been studied for reactive inertia \cite{mondal_2021a} and Gilbert damping \cite{kamra_2018}, but remain unexplored for inertial damping.

In this work, we formulate phenomenological equations of motion for the spatially uniform magnetization dynamics of a multisublattice magnet. This includes reactive inertia, Gilbert damping, and inertial damping, each with both intra- and intersublattice contributions. We then apply these equations of motion to a two-sublattice uniaxial antiferromagnet in a static magnetic field along the easy axis. We derive the linear response to a weak time-dependent magnetic field probe across all three field-dependent equilibrium phases and obtain the corresponding susceptibilities and eigenfrequencies. Our results show how reactive inertia, Gilbert damping, and inertial damping affect the frequencies and linewidths of all modes across the three magnetic phases. Based on this, we propose an experimental protocol for determining all dynamical coefficients. This article is organized as follows. In section~\ref{sec:phenomenological_dynamics} we introduce the equations of motion, in section~\ref{sec:two_sublattice_antiferromagnet} we apply them to an antiferromagnet within linear response, and in section~\ref{sec:discussion} we discuss our results.
\vspace{-0.1cm}
\section{Phenomenological dynamics} \label{sec:phenomenological_dynamics}
We describe the spatially uniform magnetization dynamics of a magnet using the sublattice magnetization vectors \eq{\bm{M}_\ell(t)}, each with fixed magnitude \eq{M_\ell = \abs{\bm{M}_\ell(t)}}. It is convenient to introduce the unit vectors \eq{\bm{m}_\ell(t) = \bm{M}_\ell(t) / M_\ell} and classify their dynamics according to time-reversal parity. Time-reversal-even terms represent reactive dynamics and do not contribute to energy dissipation, while time-reversal-odd terms may contribute to energy dissipation. We later show that the odd contributions produce damping for suitable parameter choices and therefore refer to the even and odd contributions as reactive and dissipative terms, respectively.
\vspace{-0.8cm}
\subsection{Constrained dynamics}
The reactive magnetization dynamics can be obtained from an action principle with a Lagrangian of the form
\begin{equation} \label{eq:lagrangian}
    L = L_\mr{B}\!\left(\{\bm{m}_\ell, \dot{\bm{m}}_\ell\} \right) - F\!\left(\{\bm{m}_\ell\}\right) + L_\mr{I}(\{\dot{\bm{m}}_\ell\}) ,
\end{equation}
subject to the constraint \eq{|\bm{m}_\ell| = 1}. The Berry-phase term \eq{L_\mr{B}} accounts for the geometric phase associated with the paths traced by the magnetization vectors \cite{berry_1984}, and the free energy \eq{F} encodes the magnetic interactions (exchange, anisotropy, Zeeman, and dipolar). In addition, we allow for inertial (mass-like) contributions in \eq{L_\mr{I}} that are quadratic in \eq{\dot{\bm{m}}_\ell}. In Appendix~\ref{app:lagrangian}, we give the explicit form of \eq{L} and compute the equations of motion. Here, we take an alternative approach and directly parameterize allowed contributions to the equations of motion.

The equations of motion obtained from Eq.~\eqref{eq:lagrangian} with dissipation included phenomenologically can be written as
\begin{equation} \label{eq:EOM}
    \dot{\bm{m}}_\ell = \bm{\tau}^\mr{r}_\ell + \bm{\tau}^\mr{d}_\ell.
\end{equation}
The left-hand side originates from the Berry-phase term, while the right-hand side collects all remaining contributions separated into reactive (\eq{\mr{r}}) and dissipative (\eq{\mr{d}}) terms. The reactive contributions \eq{\bm{\tau}^\mr{r}_\ell} follow from the remaining terms in the Lagrangian in Eq.~\eqref{eq:lagrangian}, whereas the dissipative contributions \eq{\bm{\tau}^\mr{d}_\ell} are introduced phenomenologically. Importantly, Eq.~\eqref{eq:EOM} is subject to the constraint \eq{\abs{\bm{m}_\ell}^2 = 1}, whose time derivative gives
\begin{equation} \label{eq:torque}
    \bm{m}_\ell \vdot \dot{\bm{m}}_\ell = 0.
\end{equation}
It follows from inserting Eq.~\eqref{eq:EOM} into Eq.~\eqref{eq:torque} that \eq{\bm{\tau}^\mr{r}_\ell} and \eq{\bm{\tau}^\mr{d}_\ell} must lie in the tangent plane to the unit sphere at \eq{\bm{m}_\ell} and are interpreted as torques acting on \eq{\bm{m}_\ell}. We note that any vector in this tangent plane can be expressed in terms of an auxiliary vector \eq{\bm{x}} through the two components \eq{\bm{m}_\ell \cross \bm{x}} and \eq{\bm{m}_\ell \cross (\bm{m}_\ell \cross \bm{x})}, provided \eq{\bm{m}_\ell \cross \bm{x} \neq \bm{0}}. These two components form orthogonal directions within the tangent plane and thus provide a convenient basis for decomposing torque contributions appearing in the equations of motion. Moreover, since \eq{\bm{m}_\ell} is odd under time reversal, the two components have opposite time-reversal parity and help us distinguish reactive from dissipative contributions.

\subsection{Reactive and dissipative torques}
We now parameterize the reactive torque contributions arising from the free energy and the inertial contributions in Eq.~\eqref{eq:lagrangian}. The free energy contributes a torque proportional to the effective field defined via \eq{\mu_0 M_\ell \bm{H}^\mr{eff}_\ell = -\partial F / \partial \bm{m}_\ell} where \eq{\mu_0} is the vacuum permeability, while the inertial term gives rise to a torque involving \eq{\ddot{\bm{m}}_\ell}. Since both \eq{\bm{H}^\mr{eff}_\ell} and \eq{\ddot{\bm{m}}_\ell} are odd under time reversal, we use the single cross product to obtain the time-reversal-even reactive torque,
\begin{equation} \label{eq:reactive_torque}
    \bm{\tau}^\mr{r}_\ell =  -\mu_0 \gamma_\ell \bm{m}_\ell \cross \bm{H}^\text{eff}_\ell + \bm{m}_\ell \cross \sum_{\ell'} \eta_{\ell \ell'} \ddot{\bm{m}}_{\ell'}.
\end{equation}
Here, \eq{\gamma_\ell > 0} is the gyromagnetic ratio that sets the strength of the effective-field precessional torque. We include only intrasublattice terms in this contribution since the effective field already incorporates intersublattice couplings through the free energy. However, for the inertial contribution parameterized by the matrix \eq{\eta}, we allow both intra- and intersublattice terms as represented by the diagonal and off-diagonal elements.

We next parameterize the dissipative torque. Since the reactive dynamics include inertial contributions, we also retain dissipative terms up to second order in time derivatives. We therefore consider dissipative contributions proportional to \eq{\dot{\bm{m}}_{\ell}} and \eq{\ddot{\bm{m}}_{\ell}}, neglecting nonlinear combinations of these since they do not contribute to the linearized magnetization dynamics. The two contributions have opposite time-reversal parity, so constructing an overall time-reversal-odd torque requires combining them with the single and double cross products, respectively. We thus write
\begin{equation} \label{eq:dissipative_torque}
    \bm{\tau}^\mr{d}_\ell = \bm{m}_\ell \cross \sum_{\ell'} \alpha_{\ell \ell'} \dot{\bm{m}}_{\ell'} + \bm{m}_\ell \cross \Big[\bm{m}_\ell \cross \sum_{\ell'} (\beta \eta)_{\ell \ell'} \ddot{\bm{m}}_{\ell'} \Big].
\end{equation}
The first and second terms describe Gilbert and inertial damping, parameterized by the dimensionless matrices \eq{\alpha} and \eq{\beta}, respectively. Both terms allow for intra- and intersublattice contributions through the diagonal and off-diagonal matrix elements. The product \eq{\beta \eta} denotes matrix multiplication, and this form is chosen to separate out the overall inertial scale set by \eq{\eta}.

\subsection{Equations of motion}
Inserting Eqs.~(\ref{eq:reactive_torque}, \ref{eq:dissipative_torque}) into Eq.~\eqref{eq:EOM} gives
\begin{align} \label{eq:finalized_eom}
    &\dot{\bm{m}}_\ell = -\gamma_\ell \mu_0 \bm{m}_\ell \cross \bm{H}^\mr{eff}_\ell + \bm{m}_\ell \cross \sum_{\ell'}\alpha_{\ell \ell'} \dot{\bm{m}}_{\ell'} 
    \\[-0.1cm]
    &+ \bm{m}_\ell \cross \sum_{\ell' } \eta_{\ell \ell'} \ddot{\bm{m}}_{\ell'} + \bm{m}_\ell \cross \Big[\bm{m}_\ell \cross \sum_{\ell'} (\beta \eta)_{\ell \ell'} \ddot{\bm{m}}_{\ell'} \Big]. \notag
\end{align}
The first term generates precessional motion of the sublattice magnetization around the effective field, and the second term provides Gilbert damping of this motion~\cite{landau_1935,kittel_1951,keffer_1952,gilbert_2004}. The third term introduces an additional timescale through second-order time derivatives and gives rise to a high-frequency nutational component~\cite{ciornei_2011, mondal_2023}. Together, these three terms constitute what is commonly referred to as the inertial Landau–Lifshitz–Gilbert equation. The last term represents the additional inertial damping contribution, which we expect to contribute primarily to the relaxation of the high-frequency nutational motion, as previously found for a ferromagnet in Ref.~\cite{giil_2026}. In the following section, we use these equations to determine the dynamical magnetic susceptibility, resonance frequencies, and linewidths of the precessional and nutational modes in a two-sublattice uniaxial antiferromagnet across all magnetic phases.
\clearpage

\section{Two-sublattice antiferromagnet} \label{sec:two_sublattice_antiferromagnet}
We consider a two-sublattice uniaxial antiferromagnet with equal sublattice saturation magnetizations, \eq{M_1 = M_2 := M}, and free energy density
\begin{align} \label{eq:free_energy}
    F(\bm{m}_1, \bm{m}_2) &= J \bm{m}_1 \vdot \bm{m}_2 - \frac{K_z}{2} [(\bm{m}_1 \vdot \hat{\bm{z}})^2 + (\bm{m}_2 \vdot \hat{\bm{z}})^2] \notag
    \\
    &-\mu_0 M \bm{H} \vdot (\bm{m}_1 + \bm{m}_2).
\end{align}
The exchange constant \eq{J > 0} favors antiparallel alignment, \eq{K_z > 0} sets the easy-axis anisotropy along \eq{\bm{z}}, and \eq{\bm{H}} is the applied magnetic field. In equilibrium, we consider a static magnetic field along the easy axis: \eq{\bm{H} = H_z \hat{\bm{z}}} with \eq{H_z > 0}. Since the free energy is invariant under interchange of the two sublattices, it is convenient to introduce the N{\'e}el and magnetization vectors
\begin{align}
    \nvec &= (\bm{m}_1 - \bm{m}_2) / 2,
    &
    \bm{m} &= (\bm{m}_1 + \bm{m}_2) / 2.
\end{align}
The N{\'e}el vector \eq{\bm{n}} describes the staggered antiferromagnetic order, while the magnetization vector \eq{\bm{m}} represents the net magnetization. The fixed-length constraint \eq{\abs{\bm{m}_1}^2 = \abs{\bm{m}_2}^2 = 1} becomes \eq{\bm{n} \vdot \bm{m} = 0} and \eq{\abs{\bm{n}}^2 + \abs{\bm{m}}^2 = 1}. Thus, the N{\'e}el and magnetization vectors are mutually orthogonal, while their magnitudes are constrained such that an increase in one is accompanied by a reduction in the other.

\subsection{Sublattice-symmetric equations of motion}
The number of independent dynamical coefficients appearing in Eq.~\eqref{eq:finalized_eom} is reduced by the assumed equivalence of the two sublattices. We take the gyromagnetic ratio to be identical on both sublattices, \eq{\gamma_1 = \gamma_2 := \gamma}, and for any matrix \eq{P \in \{\alpha, \beta, \eta \}} we define the diagonal (\eq{d}) and off-diagonal elements (\eq{x}):
\begin{align} \label{eq:intra_inter_coefficients}
    P_{11} &= P_{22} := P_d,
    &
    P_{12} &= P_{21} := P_x.
\end{align}
Rewriting Eq.~\eqref{eq:finalized_eom} in terms of the N{\'e}el and magnetization vectors and defining \eq{P_n = P_d - P_x} and \eq{P_m = P_d + P_x} yields
\begin{align} \label{eq:n_eom}
    &\dot{\nvec} = -\gamma \mu_0 \left(\nvec \cross \bm{H}_m^\mr{eff} + \bm{m} \cross \bm{H}_{\idx{n}}^\mr{eff} \right)
    \\
    +& \nvec \cross \left(\alpha_m \dot{\bm{m}} + \eta_m \ddot{\bm{m}} + \beta_{\idx{n}} \eta_{\idx{n}} \nvec \cross \ddot{\nvec} + \beta_m \eta_m \bm{m} \cross \ddot{\bm{m}} \right) \notag
    \\
    +& \bm{m} \cross \left(\alpha_{\idx{n}} \dot{\nvec} + \eta_{\idx{n}} \ddot{\nvec} + \beta_m \eta_m \nvec \cross \ddot{\bm{m}} + \beta_{\idx{n}} \eta_{\idx{n}} \bm{m} \cross \ddot{\nvec} \right), \notag
    \\[0.2cm]
    &\dot{\bm{m}} = -\gamma \mu_0 \left(\nvec \cross \bm{H}_{\idx{n}}^\mr{eff} + \bm{m} \cross \bm{H}^\mr{eff}_m \right) \label{eq:m_eom}
    \\
    +& \nvec \cross \left(\alpha_{\idx{n}} \dot{\nvec} + \eta_{\idx{n}} \ddot{\nvec} + \beta_{\idx{n}} \eta_{\idx{n}} \bm{m} \cross \ddot{\nvec} + \beta_m \eta_m \nvec \cross \ddot{\bm{m}} \right) \notag
    \\
    +& \bm{m} \cross \left(\alpha_m \dot{\bm{m}} + \eta_m \ddot{\bm{m}} + \beta_m \eta_m \bm{m} \cross \ddot{\bm{m}} + \beta_{\idx{n}} \eta_{\idx{n}} \nvec \cross \ddot{\nvec} \right), \notag
\end{align}
where
\begin{align}
    \bm{H}_n^\mr{eff} &= \frac{1}{2}(\bm{H}_1^\mr{eff} - \bm{H}_2^\mr{eff}),
    &
    \bm{H}_m^\mr{eff} &= \frac{1}{2}(\bm{H}_1^\mr{eff} + \bm{H}_2^\mr{eff}).
\end{align}
We next determine the equilibrium configurations and then linearize the equations of motion around each configuration.

\subsection{Equilibrium configurations}
Minimizing the free energy yields the three standard equilibrium configurations: a collinear antiparallel (AP) state, a noncollinear canted (CA) state, and a collinear parallel field-polarized (FP) state \cite{rezende_2019}. The phase boundaries are conveniently expressed in terms of the exchange field \eq{H_E = J /(\mu_0 M)} and anisotropy field \eq{H_A = K_z / (\mu_0 M)} through the spin-flop (SF) and saturation (sat) fields \cite{rezende_2019}:
\begin{align}
    H_\mr{SF} &= \sqrt{H_A(2H_E-H_A)},
    \\
    H_\mr{sat} &= 2H_E - H_A.
\end{align}
We assume the exchange field dominates so that \eq{H_\mr{sat} > H_\mr{SF} > 0}. The equilibrium phase can be tuned by the static magnetic field: the AP configuration is realized for \eq{H_z < H_\mr{SF}}, the CA configuration for \eq{H_\mr{SF} < H_z < H_\mr{sat}}, and the FP configuration for \eq{H_z > H_\mr{sat}}. The corresponding equilibrium configurations of the N{\'e}el and magnetization vectors are
\begin{align}
    \bm n_{(0)}^\mr{AP} &= \hat{\bm z}, & \bm m_{(0)}^\mr{AP} &= 0, \label{eq:eq_AP}
    \\
    \bm n_{(0)}^\mr{CA} &= \hat{\bm x}\sqrt{1-(H_z / H_\mr{sat})^2} , & \bm m_{(0)}^\mr{CA} &= \hat{\bm{z}} H_z / H_\mr{sat}, \label{eq:eq_CA}
    \\
    \bm n_{(0)}^\mr{FP} &= 0, & \bm m_{(0)}^\mr{FP} &= \hat{\bm z}. \label{eq:eq_FP}
\end{align}
Next, we investigate the linear response of the system to a time-dependent magnetic field probe.

\subsection{Linear response}
We probe the system with a weak time-dependent magnetic field,
\begin{equation}
    \bm{H}(t) = H_z \hat{\bm{z}} + \delta \bm{H}(t),
\end{equation}
and expand the N{\'e}el and magnetization vectors to linear order in fluctuations as
\begin{align}
    \bm{n}(t) &= \bm{n}_{(0)} + \delta \bm{n}(t),
    &
    \bm{m}(t) &= \bm{m}_{(0)} + \delta \bm{m}(t).
\end{align}
Linearizing the fixed-length constraint gives \eq{\bm n_{(0)} \vdot \delta \bm m + \bm m_{(0)} \vdot \delta \bm n = 0} and \eq{\bm n_{(0)} \vdot \delta \bm n + \bm m_{(0)} \vdot \delta \bm m = 0}. These relations restrict the allowed fluctuations about a given equilibrium configuration and reduce the maximum number of independent fluctuation components from six to four. Retaining terms to linear order in the equations of motion then yields the linearized equations of motion for each equilibrium configuration, which are given in Appendix~\ref{app:linearized_eom}.

\subsubsection{Characteristic frequency scales and approximations}
We now obtain the eigenfrequencies from the linearized equations of motion. For bright modes, the eigenfrequencies are also the resonance frequencies of the corresponding magnetic susceptibilities. Dark modes do not couple to the magnetic field at linear order, but remain part of the full mode spectrum and may be excited by other probes. The eigenspectrum is governed by four characteristic frequency scales of the antiferromagnet: the exchange frequency \eq{\omega_E = \gamma \mu_0 H_E}, the anisotropy frequency \eq{\omega_A = \gamma \mu_0 H_A}, the Zeeman frequency \eq{\omega_H = \gamma \mu_0 H_z}, and a nutational frequency \eq{\omega_\eta}.  

\begin{figure*}[p]
\centering
\begin{minipage}{\textwidth}
\vspace{-0.5cm}
\centering
\captionof{table}{Approximate eigenfrequencies to linear order in damping for well-separated low- and high-frequency modes in the exchange-dominated regime.}
\label{tab:eigenfrequencies}
\renewcommand{\arraystretch}{1.55}
\begin{ruledtabular}
\begin{tabular}{l}
\multicolumn{1}{l}{Antiparallel phase:}\\[-0.1cm]
\eq{\displaystyle \Omega^\mr{AP}_1 \approx \sqrt{2\omega_E \omega_A} -\omega_H - i \omega_E(\alpha_d - \alpha_x) \left(1 - \frac{\omega_H}{\sqrt{2\omega_E \omega_A}} \right)},
\\[0.3cm]
\eq{\displaystyle \Omega^\mr{AP}_2 \approx  \sqrt{2\omega_E \omega_A} + \omega_H - i \omega_E(\alpha_d - \alpha_x) \left(1 + \frac{\omega_H}{\sqrt{2\omega_E \omega_A}} \right)} , 
\\[0.3cm]
\eq{\displaystyle \Omega^\mr{AP}_3 \approx \Omega^\mr{AP}_4 \approx \frac{1}{\sqrt{\eta^2_d - \eta^2_x}} - i\left[ \frac{\eta_d(\alpha_d + \beta_d) - \eta_x(\alpha_x - \beta_x)}{\eta^2_d - \eta^2_x} \right]},
\\[0.3cm]
\hline
\multicolumn{1}{l}{Canted phase:}\\[-0.1cm]
\eq{\displaystyle \Omega^\mr{CA}_{1, \mr{sta}} = 0}, \hspace{0.5cm} \eq{\displaystyle\Omega^\mr{CA}_{1, \mr{rel}} \approx -i2\omega_E(\alpha_d - \alpha_x) \left(1 - \frac{\omega^2_H}{4\omega^2_E} \right)}, 
\\[0.3cm]
\eq{\displaystyle \Omega^\mr{CA}_2 \approx \sqrt{\omega^2_H - 2\omega_E \omega_A \left(1 - \frac{\omega^2_H}{4\omega^2_E} \right)} -i\omega_E \left[(\alpha_d - \alpha_x)\left(1 - \frac{\omega^2_H}{4\omega^2_E}\right) +2(\alpha_d + \alpha_x) \frac{\omega^2_H}{4\omega^2_E} \right]},
\\[0.3cm]
\eq{\displaystyle \Omega^\mr{CA}_3 \approx \frac{1}{\sqrt{\eta^2_d - \eta^2_x + 2\eta_x(\eta_d + \eta_x) \frac{\omega^2_H}{4\omega^2_E}}} - i\left\{\frac{\left[\eta_d(\alpha_d + \beta_d) - \eta_x(\alpha_x \!-\! \beta_x)\right]\left(1 - \frac{\omega^2_H}{4\omega^2_E} \right) + (\alpha_d + \beta_d + \alpha_x + \beta_x)(\eta_d + \eta_x) \frac{\omega^2_H}{4\omega^2_E}}{\eta^2_d - \eta^2_x + 2\eta_x(\eta_d + \eta_x) \frac{\omega^2_H}{4\omega^2_E}} \right\}},
\\[0.3cm]
\eq{\displaystyle \Omega^\mr{CA}_4 \approx \frac{1}{\sqrt{\eta^2_d - \eta^2_x - 2\eta_x(\eta_d - \eta_x)\frac{\omega^2_H}{4\omega^2_E}}} - i \left\{ \frac{\left[\eta_d(\alpha_d + \beta_d) - \eta_x(\alpha_x \!-\! \beta_x)\right]\left(1 - \frac{\omega^2_H}{4\omega^2_E} \right) + (\alpha_d + \beta_d - \alpha_x - \beta_x)(\eta_d - \eta_x) \frac{\omega^2_H}{4\omega^2_E}}{\eta^2_d - \eta^2_x - 2\eta_x(\eta_d - \eta_x) \frac{\omega^2_H}{4\omega^2_E}} \right\}}.
\\[0.3cm]
\hline
\multicolumn{1}{l}{Field-polarized phase:}\\[-0.1cm]
\eq{\displaystyle \Omega^\mr{FP}_1 \approx \omega_H - 2\omega_E -i(\omega_H - 2\omega_E)(\alpha_d - \alpha_x)},
\\[0.2cm]
\eq{\displaystyle \Omega^\mr{FP}_2 \approx \omega_H -i\omega_H (\alpha_d + \alpha_x)},
\\[0.2cm]
\eq{\displaystyle \Omega^\mr{FP}_3 \approx \frac{1}{\eta_d + \eta_x} -i\left( \frac{\alpha_d + \beta_d + \alpha_x + \beta_x}{\eta_d + \eta_x} \right)}, 
\\[0.3cm]
\eq{\displaystyle \Omega^\mr{FP}_4 \approx \frac{1}{\eta_d - \eta_x} -i \left( \frac{\alpha_d + \beta_d - \alpha_x - \beta_x}{\eta_d - \eta_x}\right)}.
\\[0.3cm]
\end{tabular}
\end{ruledtabular}
\end{minipage}
\begin{minipage}{\textwidth}
\includegraphics[width=\textwidth]{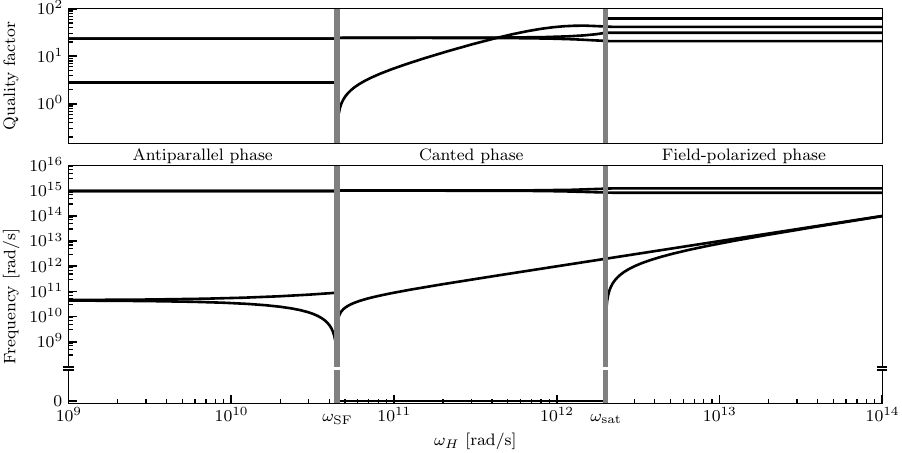}
\vspace{-0.7cm}
\caption{Eigenfrequencies across the three equilibrium phases as a function of the static magnetic field, showing the quality factors (upper panel) and the real part of the eigenfrequencies (lower panel). The parameters used are \eq{\omega_E = \SI{e12}{\radian\per\second}}, \eq{\omega_A = \SI{e9}{\radian\per\second}}, \eq{\eta_d = \SI{e-15}{\second}}, \eq{\alpha_d = 0.01}, \eq{\beta_d = 0.01}, with the cross-sublattice parameters set to \eq{p_x = 0.2 p_d} for \eq{p \in \{\eta, \alpha, \beta\}}.}
\label{fig:table_and_eigenfrequencies}
\end{minipage}
\end{figure*}

In conventional bulk antiferromagnets, the exchange frequency typically lies in the terahertz regime, whereas the anisotropy frequency is several orders of magnitude smaller \cite{baltz_2018}. The Zeeman frequency determines the equilibrium phase through its relation to the spin-flop frequency \eq{\omega_\mr{SF} = \gamma \mu_0 H_\mr{SF}} and the saturation frequency \eq{\omega_\mr{sat} = \gamma \mu_0 H_\mr{sat}}. The antiferromagnetic nutational frequency \eq{\omega_\eta} has not yet been firmly established experimentally. However, theoretical studies use nutational frequencies spanning approximately \eq{10^{12}-10^{15} \si{\radian\per\second}}~\cite{mondal_2021,mondal_2021a,mondal_2022,winter_2022,angster_2025}. We derive the dynamical magnetic susceptibilities in each equilibrium phase in Appendix~\ref{app:linearized_eom}. However, obtaining the complete eigenspectrum in closed form leads to cumbersome expressions. We therefore expand the eigenfrequencies to linear order in the damping parameters as detailed in Appendix~\ref{app:eigenspectrum}. In the main text, we further restrict attention to the approximate expressions relevant in the exchange-dominated regime, \eq{\omega_E \gg \omega_A}, and in the parameter regime where the nutational frequencies are well separated from the remaining resonances. The resulting analytical expressions simplify under these conditions and are presented in Table~\ref{tab:eigenfrequencies}.

\subsubsection{Mode structure}
Each eigenfrequency in Table~\ref{tab:eigenfrequencies} is of the form \eq{\Omega = \omega - i \Gamma}, where we refer to \eq{\omega = \Re \Omega} as the frequency and \eq{\Gamma = -\Im \Omega} as the damping rate. Figure~\ref{fig:table_and_eigenfrequencies} shows their field dependence, with the lower panel showing the frequencies \eq{\omega} and the upper panel showing the quality factors \eq{Q = \omega/(2\Gamma)}. In the exchange-dominated regime, the phase-boundary frequencies reduce to \eq{\omega_\mr{SF} \approx \sqrt{2\omega_E \omega_A}} and \eq{\omega_\mr{sat} \approx 2\omega_E}. These approximate phase boundaries should be used when interpreting the approximate eigenfrequencies in Table~\ref{tab:eigenfrequencies}. 

\paragraph{Antiparallel phase.}
The eigenfrequencies \eq{\Omega^\mr{AP}_1} and \eq{\Omega^\mr{AP}_2} correspond to the two standard circularly polarized antiferromagnetic precession modes \cite{kittel_1951}. Their frequencies are degenerate at zero magnetic field at the conventional antiferromagnetic resonance frequency. A finite magnetic field shifts the two helicities in opposite directions. Their damping
rates are governed by the exchange-enhanced difference between
the intra- and intersublattice Gilbert coefficients, consistent
with the Gilbert damping structure obtained in Ref.~\cite{kamra_2018}. The eigenfrequencies \eq{\Omega^\mr{AP}_3} and \eq{\Omega^\mr{AP}_4} correspond to approximately degenerate nutational modes. Their frequency is set by the difference of the squared intra- and intersublattice inertial coefficients, in agreement with Ref.~\cite{mondal_2021a}. Their damping rates depend on both the Gilbert and inertial damping coefficients.

\paragraph{Canted phase.}
In the CA phase, the zero-frequency Goldstone mode \eq{\Omega^\mr{CA}_{1, \mr{sta}}} reflects the freedom to choose the orientation of the spin-flop plane. We also find a purely relaxational mode \eq{\Omega^\mr{CA}_{1, \mr{rel}}}, whose damping rate is governed by the difference between the intra- and intersublattice Gilbert coefficients and decreases as the field approaches saturation. The eigenfrequency \eq{\Omega^\mr{CA}_2} describes the gapped spin-flop resonance, with frequency in agreement with Ref.~\cite{cham_2022} in the uniaxial exchange-dominated limit. Its decay rate is controlled mainly by the difference between the intra- and intersublattice Gilbert coefficients at low canting, and by their sum near saturation. The crossover is governed by the reduced field \eq{\omega_H / \omega_\mr{sat}}. Sufficiently close to the spin-flop boundary, the weak-damping approximation may break down for this mode (see Appendix~\ref{app:eigenfrequencies_CA}), limiting the validity of our expression for this eigenfrequency in this region. The eigenfrequencies \eq{\Omega^\mr{CA}_3} and \eq{\Omega^\mr{CA}_4} correspond to the nutational modes. They are nearly degenerate near the spin-flop boundary, with a frequency set by the difference of the squared intra- and intersublattice inertial coefficients. Increasing the field lifts this degeneracy. Near saturation, their frequencies are instead determined separately by the sum and difference of the two inertial coefficients. Their damping rates depend on both the Gilbert and inertial damping coefficients and evolve from a common value near the spin-flop boundary to distinct sum and difference combinations of the intra- and intersublattice damping coefficients near saturation.

\paragraph{Field-polarized phase.}
In the FP phase, the linearized dynamics separate into independent N{\'e}el and magnetization-vector sectors. The spatially uniform magnetic field probe couples only to the magnetization-vector sector at linear order. This sector is therefore bright, while the N{\'e}el-vector sector remains dark to the probe. The eigenfrequency \eq{\Omega^\mr{FP}_1} corresponds to the low-frequency N{\'e}el-vector mode and softens at the saturation boundary. Its damping rate is governed by the difference between the intra- and intersublattice Gilbert coefficients. The eigenfrequency \eq{\Omega^\mr{FP}_2} corresponds to the magnetization-vector precession mode and remains finite throughout the FP phase. Its damping rate is instead governed by the sum of the intra- and intersublattice Gilbert coefficients. The frequencies of the two low-frequency modes agree with the noninertial exchange-dominated limit of the frequencies obtained in Ref.~\cite{titov_2024}. The modes associated with \eq{\Omega^\mr{FP}_3} and \eq{\Omega^\mr{FP}_4} are the nutational modes, with \eq{\Omega^\mr{FP}_3} being bright and \eq{\Omega^\mr{FP}_4} being dark to the probe. Their frequencies are set by the sum and difference of the intra- and intersublattice inertial coefficients, respectively, while the damping rates depend on the corresponding combinations of the Gilbert and inertial damping coefficients.

\section{Discussion} \label{sec:discussion}
We first formulated phenomenological equations of motion for the spatially uniform magnetization dynamics of a general multisublattice magnet. The equations include intra- and intersublattice contributions to Gilbert damping and reactive inertia. In addition, we incorporated the recently proposed inertial damping term \cite{giil_2026}, and generalized it to include both intra- and intersublattice contributions. Like the reactive inertial term, the inertial damping term involves second-order time derivatives and should therefore generally be included when retaining contributions at that order. The resulting equations of motion in Eq.~\eqref{eq:finalized_eom} are therefore particularly relevant for ultrafast magnetization dynamics where inertial effects become important. We introduced all dynamical coefficients phenomenologically, and throughout we restricted the coefficients to the dynamically stable regime in which the frequencies are real and all damping rates are non-negative. Microscopic theories have related magnetic inertia to angular-momentum relaxation, relativistic spin-orbit effects, and retarded coupling to electronic or lattice degrees of freedom \cite{wegrowe_2012,mondal_2017,mondal_2018,quarenta_2024,johnsen_2025}. Such approaches may also clarify the microscopic origin of inertial damping and determine how the different coefficients depend on the material, temperature, and frequency. 

We next applied our phenomenological equations of motion to study the linear response of a uniaxial two-sublattice antiferromagnet in an easy-axis field to a time-dependent magnetic field probe. We obtained the dynamical susceptibilities in the three equilibrium phases. From their poles, we determined the frequencies together with their damping rates. Here we focused on the exchange-dominated regime with weak damping and a clear separation between the low- and high-frequency modes. This regime is expected for strongly exchange-coupled antiferromagnets with weak dissipation, although the frequency separation ultimately depends on the material-specific inertial timescale. We argue that this separation is experimentally motivated by conventional antiferromagnetic resonance measurements~\cite{rezende_2019}, which have generally been described without accounting for inertia, suggesting that the nutational modes lie outside the frequency range typically accessed in such experiments.

We obtained approximate expressions for all eigenfrequencies, which are summarized in Table~\ref{tab:eigenfrequencies}. The reactive inertial coefficients primarily set the nutational frequencies. Gilbert damping governs the leading low-frequency decay rates, whereas both Gilbert and inertial damping contribute to the high-frequency nutational decay. The additional inertial damping can therefore broaden the nutational resonances and affect whether they remain spectrally resolvable. This is relevant to recent terahertz measurements of inertial spin dynamics, which revealed strongly damped nutational oscillations with a broader spectral response than expected from Gilbert damping alone \cite{neeraj_2021,unikandanunni_2022}. Our eigenfrequency expressions further suggest a minimal experimental protocol for extracting the six dynamical coefficients: the intra- and intersublattice Gilbert damping coefficients, reactive inertial coefficients, and inertial damping coefficients. In principle, measurements in the AP and CA phases are sufficient. The precessional damping rates in the AP phase and their field dependence across the CA phase determine the two Gilbert coefficients. The nearly degenerate nutational frequencies in the AP phase and their splitting throughout the CA phase determine the two reactive inertial coefficients. The nutational damping rates can then be used to extract the two inertial damping coefficients. This requires frequency and linewidth measurements over a field range spanning both the AP and CA phases. Alternatively, the FP phase separates the relevant sum and difference combinations into bright magnetization-vector and dark N{\'e}el-vector modes, but this approach requires a staggered or spatially nonuniform probe to access the dark sector. 
\vspace{-0.3cm}
\section{Acknowledgements}
We acknowledge support from the Research Council of Norway through Grant No. 262633, ``Center of Excellence on Quantum Spintronics''.

\appendix

\section{Reactive equations of motion} \label{app:lagrangian}
In this appendix, we derive the reactive part of the equations of motion. We consider the Lagrangian
\begin{equation}
\begin{split}
    L &= \sum_\ell \frac{M_\ell}{\gamma_\ell} \bm{A}(\bm{m}_\ell) \vdot \dot{\bm{m}}_\ell - F(\{\bm{m}_\ell\})
    \\
    &+ \frac{1}{2} \sum_{\ell \ell'} I_{\ell \ell'} \dot{\bm{m}}_\ell \vdot \dot{\bm{m}}_{\ell '} + \sum_\ell \frac{\lambda_\ell}{2} (\bm{m}_\ell \vdot \bm{m}_\ell - 1),
\end{split}
\end{equation}
where \eq{\bm{A}} is the spin Berry-phase gauge potential satisfying \eq{\nabla_{\bm{m}_\ell} \cross \bm{A} = \bm{m}_\ell} on \eq{\abs{\bm{m}_\ell} = 1} \cite{kim_2023},  \eq{F} is the free energy, \eq{I_{\ell \ell'} = I_{\ell' \ell}} parameterizes the inertial contributions, and \eq{\lambda_\ell} are Lagrange multipliers enforcing the fixed-length constraint. The two terms entering the Euler-Lagrange equations for the Cartesian components of \eq{\bm{m}_\ell} are
\begin{align}
    \dv{t} \left(\pdv{L}{\dot{m}^i_\ell} \right) &= \frac{M_\ell}{\gamma_\ell} \sum_j \pdv{A^i}{m^j_\ell} \dot{m}^j_\ell + \sum_{\ell'} I_{\ell \ell'} \ddot{m}^i_{\ell'},
    \\
    \pdv{L}{m^i_\ell} &= \frac{M_\ell}{\gamma_\ell}\sum_j \pdv{A^j}{m^i_\ell} \dot{m}^j_\ell - \pdv{F}{m^i_\ell}  + \lambda_\ell m^i_\ell.
\end{align}
Using \eq{\bm{\nabla}_{\bm{m}_\ell} \cross \bm{A} = \bm{m}_\ell}, we obtain the equations of motion
\begin{equation}
    \frac{M_\ell}{\gamma_\ell} \bm{m}_\ell \cross \dot{\bm{m}}_\ell = -\pdv{F}{\bm{m}_\ell} - \sum_{\ell'} I_{\ell \ell'} \ddot{\bm{m}}_{\ell'} + \lambda_\ell \bm{m}_\ell.
\end{equation}
The Lagrange multiplier is eliminated by taking the cross product of this equation from the left with \eq{\bm m_\ell}. Since \eq{\bm{m}_\ell \cross (\bm{m}_\ell \cross \dot{\bm{m}}_\ell) = -\dot{\bm{m}}_\ell}, we arrive at
\begin{equation}
    \dot{\bm{m}}_\ell = -\gamma_\ell \mu_0 \bm{m}_\ell \cross \bm{H}^\mr{eff}_\ell + \sum_{\ell'} \eta_{\ell \ell'} \bm{m}_\ell \cross \ddot{\bm{m}}_{\ell'}.
\end{equation}
The effective field is defined as in the main text above Eq.~\eqref{eq:reactive_torque} and \eq{\eta_{\ell \ell'} = I_{\ell \ell'} \gamma_\ell / M_{\ell}}.

\section{Linearized equations of motion} \label{app:linearized_eom}
In this appendix, we linearize the equations of motion around the three equilibrium configurations and derive the corresponding dynamical magnetic susceptibilities. We use the Fourier-transform convention
\begin{align}
    f_\omega &= \frac{1}{\sqrt{2\pi}} \int\!\!\dd t \, e^{i\omega t} f(t),
    &
    f(t) &= \frac{1}{\sqrt{2\pi}}\int\!\!\dd\omega \, e^{-i\omega t} f_\omega,
\end{align}
and work in frequency space throughout while suppressing the subscript \eq{\omega} for notational simplicity.

\subsection{Antiparallel phase} \label{app:linearized_eom_AP}
Linearizing Eqs.~(\ref{eq:n_eom}, \ref{eq:m_eom}) around the AP ground state in Eq.~\eqref{eq:eq_AP} gives
\begin{align}
    &-i\omega \delta \bm{n} = \hat{\bm{z}} \cross \left[\omega_H \delta \bm{n} + (2\omega_E + \omega_A) \delta \bm{m} - \gamma \mu_0 \delta \bm{H} \right] \notag
    \\
    &-(i\omega \alpha_m + \omega^2 \eta_m) \hat{\bm{z}} \cross \delta \bm{m} - \omega^2 \beta_n \eta_n \hat{\bm{z}} \cross (\hat{\bm{z}} \cross \delta \bm{n}),
    \\[-0.2cm]
    &-i\omega \delta \bm{m} = \hat{\bm{z}} \cross \left(\omega_A \delta \bm{n} + \omega_H \delta \bm{m} \right) \notag
    \\
    &-(i\omega \alpha_n + \omega^2 \eta_n) \hat{\bm{z}} \cross \delta \bm{n} - \omega^2 \beta_m \eta_m \hat{\bm{z}} \cross (\hat{\bm{z}} \cross \delta \bm{m}).
\end{align}
These equations decouple into two helicity sectors using the circular basis \eq{\hat{\bm{e}}_\pm = (\hat{\bm{x}} \pm i \hat{\bm{y}})/\sqrt{2}}. The linearized equations then take the form
\begin{equation}
    D^\mr{AP}_\pm \begin{pmatrix}
        \delta n_\pm \\ \delta m_\pm
    \end{pmatrix}
    = \begin{pmatrix}
        \mp i \gamma \mu_0 \\ 0
    \end{pmatrix}
    \delta H_\pm,
\end{equation}
where \eq{\delta H_\pm} denote the two helicity components of the transverse magnetic field perturbation and the matrix elements of the dynamical matrix are
\begin{align}
    (D^\mr{AP}_\pm)_{11} &= -i\omega \mp i \omega_H - \omega^2 \beta_n \eta_n,
    \\
    (D^\mr{AP}_\pm)_{12} &= \mp i(2\omega_E + \omega_A - i \omega \alpha_m - \omega^2 \eta_m),
    \\
    (D^\mr{AP}_\pm)_{21} &= \mp i(\omega_A - i\omega \alpha_n - \omega^2 \eta_n),
    \\
    (D^\mr{AP}_\pm)_{22} &= -i\omega \mp i \omega_H - \omega^2 \beta_m \eta_m.
\end{align}
Inverting the dynamical matrix gives
\begin{align}
    \chi^\mr{AP}_{\idx{n} \pm} &= \mp i \gamma \mu_0 (D^\mr{AP}_\pm)_{22} / \det D^\mr{AP}_\pm,
    \\
    \chi^\mr{AP}_{m \pm} &= \pm i \gamma \mu_0 (D^\mr{AP}_\pm)_{21} / \det D^\mr{AP}_\pm.
\end{align}
Here we defined the N{\'e}el and magnetization susceptibilities as \eq{\delta n_\pm = \chi^\mr{AP}_{n \pm} \delta H_\pm} and \eq{\delta m_\pm = \chi^\mr{AP}_{m \pm} \delta H_\pm}.

\subsection{Canted phase}
Linearizing Eqs.~(\ref{eq:n_eom}, \ref{eq:m_eom}) around the CA ground state in Eq.~\eqref{eq:eq_CA} gives
\begin{align}
    &-i\omega \delta \bm{n} = n_0 \hat{\bm{x}} \cross \left(2 \omega_E \delta \bm{m} - \omega_A \delta m_z \hat{\bm{z}} - \gamma \mu_0 \delta \bm{H} \right)
    \\
    &-(i\omega \alpha_m + \omega^2 \eta_m) n_0 \hat{\bm{x}} \cross \delta \bm{m} -(i\omega \alpha_n + \omega^2 \eta_n) m_0 \hat{\bm{z}} \cross \delta \bm{n}  \notag
    \\
    &-\omega^2 n_0 \hat{\bm{x}} \cross \left( \beta_n \eta_n n_0 \hat{\bm{x}} \cross \delta \bm{n} + \beta_m \eta_m m_0 \hat{\bm{z}} \cross \delta \bm{m} \right) \notag
    \\
    &-\omega^2 m_0 \hat{\bm{z}} \cross \left(\beta_m \eta_m n_0 \hat{\bm{x}} \cross \delta \bm{m} + \beta_n \eta_n m_0 \hat{\bm{z}} \cross \delta \bm{n} \right), \notag \label{eq:eqn_CA_lin}
    \\[0.3cm]
    &-i\omega \delta \bm{m} = -n_0 \omega_A \hat{\bm{x}}  \cross \delta n_z \hat{\bm{z}} - \gamma \mu_0 m_0 \hat{\bm{z}} \cross \delta \bm{H}
    \\
    &+ (\omega_H + \omega_A m_0) \hat{\bm{z}} \cross \delta \bm{m} \notag
    \\
    &-(i\omega \alpha_n + \omega^2 \eta_n) n_0 \hat{\bm{x}} \cross \delta \bm{n} - (i\omega \alpha_m + \omega^2 \eta_m) m_0 \hat{\bm{z}} \cross \delta \bm{m} \notag
    \\
    &-\omega^2 n_0 \hat{\bm{x}} \cross \left(\beta_n \eta_n m_0 \hat{\bm{z}} \cross \delta \bm{n} + \beta_m \eta_m n_0 \hat{\bm{x}} \cross \delta \bm{m} \right) \notag
    \\
    &-\omega^2 m_0 \hat{\bm{z}} \cross \left( \beta_m \eta_m m_0 \hat{\bm{z}} \cross \delta \bm{m} + \beta_n \eta_n n_0 \hat{\bm{x}} \cross \delta \bm{n} \right) \notag \label{eq:eqm_CA_lin}
    \end{align}
where \eq{m_0 = \omega_H / \omega_\mr{sat}} and \eq{n_0 = \sqrt{1 - m_0^2}}. The linearized fixed-length constraint restricts
\begin{align}
    \delta n_x &= - \frac{m_0}{n_0} \delta m_z, & \delta m_x &= - \frac{m_0}{n_0} \delta n_z,
\end{align}
leaving four independent variables: \eq{\delta n_y, \delta n_z, \delta m_y, \delta m_z}. Using these variables, Eqs.~(\ref{eq:eqn_CA_lin}, \ref{eq:eqm_CA_lin}) decouple into two sectors. We denote them by \eq{\perp} and \eq{\parallel} according to the orientation of the driving field as
\begin{align}
    D^\mr{CA}_\perp 
    \begin{pmatrix}
        \delta n_z
        \\ 
        \delta m_y
    \end{pmatrix}
    &=
    R^\mr{CA}_\perp
    \begin{pmatrix}
        \delta H_x 
        \\
         \delta H_y
    \end{pmatrix},
    \\
    D^\mr{CA}_\parallel 
    \begin{pmatrix}
        \delta n_y
        \\ 
        \delta m_z
    \end{pmatrix}
    &=
    R^\mr{CA}_\parallel \delta H_z.
\end{align}
The matrix elements of the dynamical matrices are
\begin{align}
    (D^\mr{CA}_\perp)_{11} &= -i\omega - \omega^2 (n^2_0 \beta_n \eta_n + m^2_0 \beta_m \eta_m),
    \\
    (D^\mr{CA}_\perp)_{12} &= -n_0 2\omega_E + n_0 A_m,
    \\
    (D^\mr{CA}_\perp)_{21} &= -n_0\omega_A + \frac{m^2_0}{n_0}2\omega_E -n_0 A_n - \frac{m^2_0}{n_0} A_m,
    \\
    (D^\mr{CA}_\perp)_{22} &= -i\omega - \omega^2 \beta_m \eta_m,
    \\
    (D^\mr{CA}_\parallel)_{11} &= -i\omega - \omega^2 \beta_n \eta_n,
    \\
    (D^\mr{CA}_\parallel)_{12} &= n_0 \omega_\mr{sat} - n_0 A_m - \frac{m^2_0}{n_0} A_n,
    \\
    (D^\mr{CA}_{\parallel})_{21} &= n_0 A_n,
    \\
    (D^\mr{CA}_\parallel)_{22} &= -i\omega - \omega^2(m^2_0 \beta_n \eta_n + n^2_0 \beta_m \eta_m),
\end{align}
where \eq{A_{n/m} = i\omega \alpha_{n/m} + \omega^2 \eta_{n/m}}, and the field-coupling matrices are
\begin{align}
    R^\mr{CA}_\perp &= \gamma \mu_0
    \begin{pmatrix}
        0 & -n_0
        \\
        -m_0 & 0
    \end{pmatrix},
    &
    R^\mr{CA}_\parallel &= \gamma \mu_0
    \begin{pmatrix}
        n_0 \\ 0
    \end{pmatrix}.
\end{align}
The dynamical susceptibilities in the two sectors are then given by
\begin{align}
    \chi^\mr{CA}_\perp &= (D^\mr{CA}_\perp)^{-1} R^\mr{CA}_\perp,
    \\
    \chi^\mr{CA}_\parallel &= (D^\mr{CA}_\parallel)^{-1} R^\mr{CA}_\parallel.
\end{align}
We next turn to the field-polarized phase, where the collinear ground state leads to simpler linearized dynamics.

\subsection{Field-polarized phase}
Linearizing Eqs.~(\ref{eq:n_eom}, \ref{eq:m_eom}) around the FP ground state in Eq.~\eqref{eq:eq_FP} gives
\begin{align}
    &-i\omega \delta \bm{n} = (\omega_H - \omega_\mr{sat}) \hat{\bm{z}} \cross \delta \bm{n}
    \\
    &-(i\omega \alpha_n + \omega^2 \eta_n) \hat{\bm{z}} \cross \delta \bm{n} - \omega^2 \beta_n \eta_n \hat{\bm{z}} \cross (\hat{\bm{z}} \cross \delta \bm{n}), \notag
    \\[0.3cm]
    &-i\omega \delta \bm{m} = \hat{\bm{z}} \cross \left[(\omega_H + \omega_A) \delta \bm{m} - \gamma \mu_0 \delta \bm{H} \right]
    \\
    &-(i\omega \alpha_m + \omega^2 \eta_m) \hat{\bm{z}} \cross \delta \bm{m} - \omega^2 \beta_m \eta_m \hat{\bm{z}} \cross (\hat{\bm{z}} \cross \delta \bm{m}). \notag
\end{align}
This shows that the linearized N{\'e}el and magnetization sectors decouple. Transforming to the circular basis as defined in Appendix~\ref{app:linearized_eom_AP} diagonalizes both sectors, and the equations become
\begin{equation}
    D^\mr{FP}_\pm
    \begin{pmatrix}
        \delta n_\pm \\ \delta m_\pm
    \end{pmatrix}
    = \begin{pmatrix}
        0 \\ \mp i\gamma \mu_0 
    \end{pmatrix}
    \delta H_\pm.
\end{equation}
The dynamical matrix \eq{D^\mr{FP}_\pm} is diagonal with elements
\begin{align}
    (D^\mr{FP}_\pm)_{11} &= \mp i (\omega_H - \omega_\mr{sat} - i\omega \alpha_n - \omega^2 \eta_n) \notag
    \\
    &\hspace{0.4cm} -i\omega - \omega^2 \beta_n \eta_n ,
    \\
    (D^\mr{FP}_\pm)_{22} &= \mp i(\omega_H + \omega_A - i\omega \alpha_m - \omega^2\eta_m) \notag 
    \\
    & \hspace{0.4cm} -i\omega - \omega^2 \beta_m \eta_m,
\end{align}
and the susceptibilities are \eq{\chi^\mr{FP}_{\idx{n} \pm} = 0} and 
\begin{align}
    \chi^\mr{FP}_{m \pm} &= \mp i\gamma \mu_0 / (D^\mr{FP}_\pm)_{22}.
\end{align}
The vanishing N{\'e}el susceptibility follows from the decoupling of the two sectors at linear order, with the probe field coupling only to the magnetization vector. 

\section{Eigenfrequencies} \label{app:eigenspectrum}
In this appendix, we determine the eigenfrequencies. They are obtained from the zeros of the determinants of the dynamical matrices, equivalently from the poles of their inverse corresponding to the susceptibilities. The determinants of the dynamical matrices are generally polynomial in \eq{\omega}, where we here find the roots perturbatively to first order in the damping parameters. To describe the perturbative procedure, we consider the polynomial \eq{P(\epsilon, \omega) = \sum_{n = 0}^\infty p_n(\omega) \epsilon^n}, where the parameter \eq{\epsilon} tracks the order in damping. We seek a root of the form \eq{\Omega = \Omega_{(0)} + \Omega_{(1)} \epsilon + \mathcal{O}(\epsilon^2)}. Expanding \eq{P(\epsilon, \Omega) = 0} to first order in \eq{\epsilon} gives
\begin{align}
	0 &= p_0\left(\Omega_{(0)}\right), \label{app:zeroth-order}
	\\
	\Omega_{(1)} &= -p_1\left(\Omega_{(0)}\right) / p'_0\left(\Omega_{(0)}\right), \label{app:first-order}
\end{align}
provided that \eq{p'_0\left(\Omega_{(0)}\right) \neq 0}. Thus, the zeroth-order contribution is obtained by solving Eq.~\eqref{app:zeroth-order}, while the first-order correction follows by using Eq.~\eqref{app:first-order}. Before computing the eigenfrequencies, we introduce two approximations motivated by the frequency scales introduced in the main text. We use the notation
\begin{align}
    a \underset{1}{\approx} b &\qq{when} \omega_A \ll \eta^{-1}_m, \eta^{-1}_n,
    \\
    a \underset{2}{\approx} b &\qq{when} \omega_E, \omega_A \ll \eta^{-1}_m, \eta^{-1}_n,
\end{align}
for two quantities \eq{a} and \eq{b}. Approximation 2 is more restrictive than approximation 1, as it additionally requires the exchange frequency \eq{\omega_E} to be small compared with the inertial frequency scales. 

\subsection{Antiparallel phase}
We determine the eigenfrequencies in the AP phase from the roots of
\begin{equation}
    \det D^\mr{AP}_\pm = a_\pm \omega^0 + b_\pm \omega^1 + c_\pm \omega^2 + d_\pm \omega^3 + e_\pm \omega^4,
\end{equation}
where
\begin{align}
    a_\pm &= \omega_A(2\omega_E + \omega_A) - \omega^2_H,
    \\
    b_\pm &= -i[(2 \omega_E + \omega_A)\alpha_n + \omega_A \alpha_m] \mp 2\omega_H,
    \\
    c_\pm &= -1 - \alpha_m \alpha_n - (2\omega_E + \omega_A) \eta_n - \omega_A \eta_m \notag 
    \\
    &\pm i\omega_H(\beta_m \eta_m + \beta_n \eta_n),
    \\
    d_\pm &= i [(\alpha_n + \beta_m)\eta_m + (\alpha_m + \beta_n)\eta_n],
    \\
    e_\pm &= (1 + \beta_m \beta_n) \eta_m \eta_n.
\end{align}
We first consider the zeroth-order contribution in damping, which satisfies
\begin{equation}
	0 = \omega_0^2 - \big[\Omega^\mr{AP}_{(0) \pm} \pm \omega_H\big]^2 -  \big[\Omega^\mr{AP}_{(0) \pm}\big]^2 f + \big[\Omega^\mr{AP}_{\pm (0)} \big]^4 \eta_m \eta_n,
\end{equation}
where \eq{\omega_0} is the conventional antiferromagnetic resonance frequency and \eq{f} characterizes the relative importance of the inertial terms on the exchange scale,
\begin{align}
    \omega_0^2 &= \omega_A (2\omega_E + \omega_A),
    \\
    f &= (2 \omega_E + \omega_A) \eta_n + \omega_A \eta_m.
\end{align}
Within approximation 1, we find that the low- and high-frequency branches are well separated. We therefore neglect the quartic term for the low-frequency branches and the constant and linear terms for the high-frequency branches. This yields
\begin{align}
	\Omega^\mr{AP}_{\pm (0) \mr{low}} &\underset{1}{\approx} \Big[ \omega_0 \sqrt{1 + f (1 - \omega^2_H / \omega^2_0)} \mp \omega_H \Big] \big/ (1 + f),
	\\
	\Omega^\mr{AP}_{\pm (0) \mr{high}} &\underset{1}{\approx} \sqrt{(1 + f) / (\eta_m \eta_n)}. 
\end{align}
Both branches depend on \eq{f}, which is not generically negligible and depends on the dimensionless products of the inertial coefficients and the exchange and anisotropy frequencies.

The first-order correction in the damping parameters is
\begin{equation}
	\Omega^\mr{AP}_{\pm (1)} = \frac{i \Im (b_\pm) \Omega^\mr{AP}_{\pm (0)} + i \Im(c_\pm) \big[\Omega^\mr{AP}_{\pm (0)} \big]^2 + d_\pm \big[\Omega^\mr{AP}_{\pm (0)} \big]^3}{2(1 + f) \Omega^\mr{AP}_{\pm (0)} \pm 2 \omega_H - 4\eta_m \eta_n \big[\Omega^\mr{AP}_{\pm (0)} \big]^3}.
\end{equation}
Inserting the approximate undamped frequencies of the two branches and retaining only the leading contributions within approximation 2, which allows the resulting expressions to be simplified, we obtain
\begin{align}
	\Omega^\mr{AP}_{\pm (1) \mr{low}} &\underset{2}{\approx} -i \left[\omega_E \alpha_n + \omega_A(\alpha_m + \alpha_n) / 2 \right] (1 \mp \omega_H / \omega_0),
	\\
	\Omega^\mr{AP}_{\pm (1) \mr{high}} &\underset{2}{\approx} - i\left[(\alpha_n + \beta_m) / \eta_n + (\alpha_m + \beta_n) / \eta_m \right] / 2.
\end{align}
We next determine the eigenfrequencies in the CA phase.

\subsection{Canted phase} \label{app:eigenfrequencies_CA}
The dynamical matrix in the CA phase is block diagonal, so that its determinant factorizes as \eq{\det D^\mr{CA} = \det D^\mr{CA}_\perp \det D^\mr{CA}_\parallel}. This allows the eigenfrequencies associated with the transverse and longitudinal sectors to be determined independently. Before doing so, it is convenient to introduce the dimensionless reduced field
\begin{equation}
    h = \omega_H / \omega_\mr{sat},
\end{equation}
which equals the equilibrium magnetization vector amplitude in the canted phase. It is generally small near the spin-flop boundary in the exchange-dominated regime and approaches unity at saturation. 

We determine the eigenfrequencies in the \eq{\perp} sector from the roots of 
\begin{equation}
    \det D^\mr{CA}_\perp = a_\perp \omega^0 + b_\perp \omega^1 + c_\perp \omega^2 + d_\perp \omega^3 + e_\perp \omega^4,
\end{equation}
where
\begin{align}
    a_\perp &= -2\omega_E \omega_A (1 - h^2) + 4\omega^2_E h^2, 
    \\
    b_\perp &= -i[(2\alpha_n \omega_E - \alpha_m \omega_A)(1 - h^2) + 4\omega_E \alpha_m h^2],
    \\
    c_\perp &= -1 + (\eta_m \omega_A - 2\eta_n \omega_E - \alpha_n \alpha_m) (1 - h^2) \notag
    \\
    &-(\alpha^2_m + 4\eta_m \omega_E) h^2,
    \\
    d_\perp &= i[(\alpha_n + \beta_m)\eta_m + (\alpha_m + \beta_n)\eta_n] (1 - h^2) \notag
    \\
    &+i2(\alpha_m + \beta_m) \eta_m h^2,
    \\
    e_\perp &= \eta_m \eta_n(1 + \beta_m \beta_n) (1 - h^2) + \eta^2_m (1 + \beta^2_m)h^2.
\end{align}
At zeroth order in the damping parameters, we find
\begin{equation}
    \big[\Omega^\mr{CA}_{\perp (0)}\big]^2 = \frac{c_\perp}{2 e_\perp} \left(-1 \pm \sqrt{1 - \frac{4 a_\perp e_\perp}{c^2_\perp}} \, \right) \eval_{\alpha = \beta = 0},
\end{equation}
where \eq{+ (-)} labels the low-frequency (high-frequency) branch. Under approximation 2, these expressions reduce to
\begin{align}
    \big[\Omega^\mr{CA}_{\perp (0) \mr{low}}\big]^2 &\underset{2}{\approx} a_\perp,
    &
    \big[\Omega^\mr{CA}_{\perp (0) \mr{high}}\big]^2 &\underset{2}{\approx} 1 / e_\perp \eval_{\beta = 0}.
\end{align}
The first-order corrections are
\begin{equation}
    \Omega^\mr{CA}_{\perp (1)} = -\frac{b_\perp + d_\perp \big[\Omega^\mr{CA}_{\perp(0)}\big]^2}{\left(2 c_\perp + 4 e_\perp \big[\Omega^\mr{CA}_{\perp(0)}\big]^2 \right) \Big|_{\alpha = \beta = 0}}.
\end{equation}
Using approximation 2, these reduce to
\begin{align}
    \Omega^\mr{CA}_{\perp (1) \mr{low}} &\underset{2}{\approx} b_\perp / 2,
    &
    \Omega^\mr{CA}_{\perp (1) \mr{high}} &\underset{2}{\approx} - d_\perp / (2 e_\perp |_{\beta = 0}).
\end{align}
We note that at the spin-flop boundary in the exchange-dominated regime, \eq{\abs\big{\Omega^\mr{CA}_{\perp(1) \mr{low}}} / \Omega^\mr{CA}_{\perp(0) \mr{low}} \approx \alpha_n \omega_E / (\sqrt{2} \omega_A)}, which is not necessarily small even for small Gilbert damping coefficients. The first-order expansion may therefore fail for this specific mode sufficiently close to the spin-flop boundary.

We next consider the \eq{\parallel} sector and determine the eigenfrequencies from the roots of
\begin{equation}
    \det D^\mr{CA}_\parallel = b_\parallel \omega^1 + c_\parallel \omega^2 + d_\parallel \omega^3 + e_\parallel \omega^4,
\end{equation}
where
\begin{align}
    b_\parallel &= -i \alpha_n \omega_\mr{sat} (1 - h^2),
    \\
    c_\parallel &= -1 -(\alpha_m \alpha_n + \eta_n \omega_\mr{sat}) (1 - h^2) - \alpha^2_n h^2,
    \\
    d_\parallel &= i[(\alpha_m + \beta_n)\eta_n + (\alpha_n + \beta_m)\eta_m](1 - h^2)  \notag
    \\
    &+ i2(\alpha_n + \beta_n) \eta_n h^2,
    \\
    e_\parallel &= \eta_m \eta_n (1 + \beta_m \beta_n) (1 - h^2) + \eta_n^2 (1 + \beta_n^2) h^2.
\end{align}
This polynomial contains an overall factor of \eq{\omega}, corresponding to an exact zero-frequency Goldstone mode. At zeroth order in damping, we find a second zero-frequency root, whose first-order correction in damping is
\begin{equation}
    \Omega^\mr{CA}_{\parallel \mr{rel}} = - \frac{b_{\parallel}}{c_{\parallel} \eval_{\alpha = 0}} = -i\frac{\alpha_n \omega_\mr{sat} (1 - h^2)}{[1 + \eta_n \omega_\mr{sat}(1-h^2)]}.
\end{equation}
This mode is purely relaxational. The other two roots form the high-frequency pair. Their zeroth- and first-order contributions are
\begin{align}
    [\Omega^\mr{CA}_{\parallel (0) \text{high}}]^2 &= -\frac{c_\parallel}{e_\parallel} \eval_{\alpha = \beta = 0} \underset{2}{\approx} \frac{1}{e_\parallel} \eval_{\beta = 0},
    \\
    \Omega^\mr{CA}_{\parallel (1) \text{high}} &= \frac{b_\parallel}{2 c_\parallel |_{\alpha = 0}} - \frac{d_\parallel}{2e_\parallel |_{\beta = 0}} \underset{2}{\approx} -\frac{d_\parallel}{2 e_\parallel |_{\beta = 0}}.
\end{align}
We next determine the eigenfrequencies in the FP phase.
\vspace{-0.7cm}
\subsection{Field-polarized phase}
The linearized equations of motion in the FP phase decoupled into separate magnetization (m) and N{\'e}el-vector (n) sectors. We first consider the magnetization sector, in which the eigenfrequencies are obtained from the zeros of \eq{(D^\mr{FP}_{\pm})_{22}}. We find
\begin{align}
    \Omega^\mr{FP}_{m\pm(0)} &= \frac{\pm 1}{2\eta_m} \left[1 + s \sqrt{1 + 4 (\omega_H + \omega_A)\eta_m } \right] \label{eq:sqrtt},
    \\
    \Omega^\mr{FP}_{m \pm(1) } &= -i \Omega^\mr{FP}_{m \pm(0)  } \left[ \frac{\pm \alpha_m + \eta_m \beta_m  \Omega^\mr{FP}_{m \pm(0) }}{\pm 2 \eta_m \Omega^\mr{FP}_{m \pm (0) } - 1} \right]. \label{eq:sqrtt2}
\end{align}
Here \eq{s \in \{-1, +1\}} labels the two solutions of the quadratic formula. Since \eq{\omega_H \geq \omega_\mr{sat} = 2 \omega_E - \omega_A} in the FP phase, we expand the square root in Eq.~\eqref{eq:sqrtt} using approximation 2 together with the field restriction \eq{\omega_H \eta_m \ll 1}. The two choices \eq{s = \pm 1} yield the low- and high-frequency branches,
\begin{align}
    \Omega^\mr{FP}_{m \pm(0)  \mr{low}} &\underset{2}{\approx} \mp(\omega_H + \omega_A), 
    \\
    \Omega^\mr{FP}_{m \pm(0)  \mr{high}} &\underset{2}{\approx} \pm 1 / \eta_m.
\end{align}
Inserting these into Eq.~\eqref{eq:sqrtt2} gives the first-order corrections in damping
\begin{align}
    \Omega^\mr{FP}_{m \pm (1) \mr{low}} &\underset{2}{\approx} -i (\omega_H + \omega_A) \alpha_m, 
    \\
    \Omega^\mr{FP}_{m \pm (1) \mr{high}} &\underset{2}{\approx} -i (\alpha_m + \beta_m) / \eta_m.
\end{align}

The eigenfrequencies associated with the N{\'e}el vector dynamics are obtained from the zeros of \eq{(D^\mr{FP}_\pm)_{11}}. We obtain
\begin{align}
    \Omega^\mr{FP}_{n\pm(0)} &= \frac{\pm1}{2\eta_n}\left[1 + s \sqrt{1 + 4(\omega_H - \omega_\mr{sat})\eta_n} \right],
    \\
    \Omega^\mr{FP}_{n \pm (1)} &= - i \Omega^\mr{FP}_{n \pm (0)} \left[\frac{\pm \alpha_n + \eta_n \beta_n \Omega^\mr{FP}_{n \pm (0)}}{\pm 2 \eta_n \Omega^\mr{FP}_{n \pm (0)} - 1} \right].
\end{align}
In the experimentally relevant regime near the phase boundary, we have \eq{(\omega_H - \omega_\mr{sat}) \eta_n \ll 1}, so that
\begin{align}
    \Omega^\mr{FP}_{n \pm (0) \mr{low}} &\approx \mp (\omega_H - \omega_\mr{sat}), 
    \\
    \Omega^\mr{FP}_{n \pm (0) \mr{high} } &\approx \pm 1 / \eta_n,
\end{align}
with the corresponding first-order corrections
\begin{align}
    \Omega^\mr{FP}_{n \pm (1) \mr{low}} &\approx -i (\omega_H - \omega_\mr{sat}) \alpha_n,  
    \\
    \Omega^\mr{FP}_{n \pm (1) \mr{high} } &\approx -i (\alpha_n + \beta_n) / \eta_n.
\end{align}
We next summarize the eigenfrequencies across all phases in the exchange-dominated well-separated regime.
\vspace{-0.2cm}
\subsection{Exchange-dominated well-separated regime}
We now reduce the phase-specific results to the forms used in Table~\ref{tab:eigenfrequencies}. We assume the exchange-dominated limit, \eq{\omega_E \gg \omega_A}, together with approximation 2, under which the low- and high-frequency sectors are well separated. For the AP phase, this gives
\begin{align}
    \Omega^\mr{AP}_{\pm \mr{low}} &\approx \sqrt{2\omega_E \omega_A} \mp \omega_H - i\omega_E \alpha_n \left(1 \mp \frac{\omega_H}{\sqrt{2\omega_E \omega_A}}\right),
    \\
    \Omega^\mr{AP}_{\pm \mr{high}} &\approx \frac{1}{\sqrt{\eta_m \eta_n}} - i\frac{[\eta_m(\alpha_n + \beta_m) + \eta_n(\alpha_m + \beta_n)]}{2\eta_m \eta_n}.
\end{align}
For the CA phase, applying the same approximations gives
\begin{align}
    \Omega^\mr{CA}_{\perp \mr{low}} &\approx \sqrt{4 \omega^2_E h^2 - 2 \omega_E \omega_A(1 - h^2)} \notag
    \\
    &-i\left[\alpha_n \omega_E(1-h^2) + 2 \omega_E \alpha_m h^2 \right],
    \\
    \Omega^\mr{CA}_{\perp \mr{high}} &\approx 1 / \sqrt{\eta_m \eta_n (1 - h^2) + \eta^2_m h^2} \notag
    \\
    &-i \Im(d_\perp) / \left[2\eta_m \eta_n (1 - h^2) + 2\eta^2_m h^2 \right],
    \\
    \Omega^\mr{CA}_{\parallel \mr{sta}} &= 0,
    \\
    \Omega^\mr{CA}_{\parallel \mr{rel}} &\approx -i\alpha_n \omega_\mr{sat}(1 - h^2),
    \\
    \Omega^\mr{CA}_{\parallel \mr{high}} &\approx 1 / \sqrt{\eta_m \eta_n (1 - h^2) + \eta^2_n h^2} \notag
    \\
    &-i \Im(d_\parallel) / \left[2\eta_m \eta_n (1 - h^2) + 2\eta^2_n h^2 \right].
\end{align}
For the FP phase, we retain only the positive frequencies and obtain
\begin{align}
    \Omega^\mr{FP}_{m - \mr{low}} &\approx \omega_H - i\omega_H \alpha_m,
    \\
    \Omega^\mr{FP}_{m + \mr{high}} &\approx 1 / \eta_m - i (\alpha_m + \beta_m) / \eta_m,
    \\
    \Omega^\mr{FP}_{n - \mr{low}} &\approx \omega_H - 2\omega_E - i(\omega_H - 2\omega_E) \alpha_n,
    \\
    \Omega^\mr{FP}_{n + \mr{high}} &\approx 1 / \eta_n - i(\alpha_n + \beta_n) / \eta_n.
\end{align}
The expressions in Table~\ref{tab:eigenfrequencies} are obtained by substituting the intra- and intersublattice combinations for the dynamical coefficients defined below Eq.~\eqref{eq:intra_inter_coefficients}. 

\bibliography{references}

@article{landau_1935,
    author  = {Landau, L. D. and Lifshitz, E. M.},
    title   = {On the Theory of the Dispersion of Magnetic Permeability in Ferromagnetic Bodies},
    journal = {Phys. Z. Sowjetunion},
    volume  = {8},
    pages   = {153--169},
    year    = {1935}
}

@article{gilbert_2004,
    author  = {Gilbert, T. L.},
    title   = {A Phenomenological Theory of Damping in Ferromagnetic Materials},
    journal = {IEEE Trans. Magn.},
    volume  = {40},
    pages   = {3443--3449},
    year    = {2004},
    doi     = {10.1109/TMAG.2004.836740}
}

@article{mondal_2023,
    author  = {Mondal, Ritwik and R{\'o}zsa, Levente and Farle, Michael and Oppeneer, Peter M. and Nowak, Ulrich and Cherkasskii, Mikhail},
    title   = {Inertial Effects in Ultrafast Spin Dynamics},
    journal = {J. Magn. Magn. Mater.},
    volume  = {579},
    pages   = {170830},
    year    = {2023},
    doi     = {10.1016/j.jmmm.2023.170830}
}

@article{ciornei_2011,
    author  = {Ciornei, M.-C. and Rub{\'i}, J. M. and Wegrowe, J.-E.},
    title   = {Magnetization Dynamics in the Inertial Regime: Nutation Predicted at Short Time Scales},
    journal = {Phys. Rev. B},
    volume  = {83},
    pages   = {020410},
    year    = {2011},
    doi     = {10.1103/PhysRevB.83.020410}
}

@article{heitzer_2024,
    author  = {Heitzer, R. C. and Pinto, F. and Rodr{\'i}guez, E. and Rodr{\'i}guez-Su{\'a}rez, R. and Maze, J. R.},
    title   = {Characterization of Antiferromagnetic Magnons Using Nitrogen Vacancy Center Relaxometry},
    journal = {Phys. Rev. B},
    volume  = {110},
    pages   = {134431},
    year    = {2024},
    doi     = {10.1103/PhysRevB.110.134431}
}

@article{fahnle_2011,
    author  = {F{\"a}hnle, Manfred and Steiauf, Daniel and Illg, Christian},
    title   = {Generalized Gilbert Equation Including Inertial Damping: Derivation from an Extended Breathing Fermi Surface Model},
    journal = {Phys. Rev. B},
    volume  = {84},
    pages   = {172403},
    year    = {2011},
    doi     = {10.1103/PhysRevB.84.172403}
}

@article{bhattacharjee_2012,
    author  = {Bhattacharjee, Satadeep and Nordstr{\"o}m, Lars and Fransson, Jonas},
    title   = {Atomistic Spin Dynamic Method with Both Damping and Moment of Inertia Effects Included from First Principles},
    journal = {Phys. Rev. Lett.},
    volume  = {108},
    pages   = {057204},
    year    = {2012},
    doi     = {10.1103/PhysRevLett.108.057204}
}

@article{giil_2026,
    author = {Giil, Hans Gløckner and Brataas, Arne},
    title = {Nutational spin pumping and dissipation},
    journal = {Phys. Rev. B},
    volume = {114},
    pages = {L140413},
    year = {2026},
    doi = {10.1103/3rrz-z4dm},
}

@article{neeraj_2021,
    author  = {Neeraj, Kumar and Awari, Nilesh and Kovalev, Sergey and Polley, Debanjan and Zhou Hagstr{\"o}m, Nanna and Arekapudi, Sri Sai Phani Kanth and Semisalova, Anna and Lenz, Kilian and Green, Bertram and Deinert, Jan-Christoph and Ilyakov, Igor and Chen, Min and Bawatna, Mohammed and Scalera, Valentino and {d'Aquino}, Massimiliano and Serpico, Claudio and Hellwig, Olav and Wegrowe, Jean-Eric and Gensch, Michael and Bonetti, Stefano},
    title   = {Inertial Spin Dynamics in Ferromagnets},
    journal = {Nat. Phys.},
    volume  = {17},
    pages   = {245--250},
    year    = {2021},
    doi     = {10.1038/s41567-020-01040-y}
}

@article{unikandanunni_2022,
    author  = {Unikandanunni, Vivek and Medapalli, Rajasekhar and Asa, Marco and Albisetti, Edoardo and Petti, Daniela and Bertacco, Riccardo and Fullerton, Eric E. and Bonetti, Stefano},
    title   = {Inertial Spin Dynamics in Epitaxial Cobalt Films},
    journal = {Phys. Rev. Lett.},
    volume  = {129},
    pages   = {237201},
    year    = {2022},
    doi     = {10.1103/PhysRevLett.129.237201}
}

@article{mondal_2021,
    author  = {Mondal, Ritwik and Gro{\ss}enbach, Sebastian and R{\'o}zsa, Levente and Nowak, Ulrich},
    title   = {Nutation in Antiferromagnetic Resonance},
    journal = {Phys. Rev. B},
    volume  = {103},
    pages   = {104404},
    year    = {2021},
    doi     = {10.1103/PhysRevB.103.104404}
}

@article{mondal_2021a,
    author  = {Mondal, Ritwik and Oppeneer, Peter M.},
    title   = {Influence of Intersublattice Coupling on the Terahertz Nutation Spin Dynamics in Antiferromagnets},
    journal = {Phys. Rev. B},
    volume  = {104},
    pages   = {104405},
    year    = {2021},
    doi     = {10.1103/PhysRevB.104.104405}
}

@article{mondal_2022,
    author  = {Mondal, Ritwik and R{\'o}zsa, Levente},
    title   = {Inertial Spin Waves in Ferromagnets and Antiferromagnets},
    journal = {Phys. Rev. B},
    volume  = {106},
    pages   = {134422},
    year    = {2022},
    doi     = {10.1103/PhysRevB.106.134422}
}

@article{winter_2022,
    author  = {Winter, Lucas and Gro{\ss}enbach, Sebastian and Nowak, Ulrich and R{\'o}zsa, Levente},
    title   = {Nutational Switching in Ferromagnets and Antiferromagnets},
    journal = {Phys. Rev. B},
    volume  = {106},
    pages   = {214403},
    year    = {2022},
    doi     = {10.1103/PhysRevB.106.214403}
}

@article{titov_2024,
    author  = {Titov, Sergei V. and Dowling, William J. and Titov, Anton S. and Fedorov, Andrey S.},
    title   = {Antiferromagnetic and Nutation Resonance Frequencies of Antiferromagnets at an Arbitrary Strength of the Applied Dc Field},
    journal = {J. Appl. Phys.},
    volume  = {135},
    pages   = {093903},
    year    = {2024},
    doi     = {10.1063/5.0196622}
}

@article{angster_2025,
    author  = {Angster, David and Dannegger, Tobias and Schlegel, Julius and Evers, Martin and Nowak, Ulrich},
    title   = {Nutational Resonance Modes in Antiferromagnetic Materials},
    journal = {Sci. Rep.},
    volume  = {15},
    pages   = {21543},
    year    = {2025},
    doi     = {10.1038/s41598-025-08746-0}
}

@article{kamra_2018,
    author  = {Kamra, Akashdeep and Troncoso, Roberto E. and Belzig, Wolfgang and Brataas, Arne},
    title   = {Gilbert Damping Phenomenology for Two-Sublattice Magnets},
    journal = {Phys. Rev. B},
    volume  = {98},
    pages   = {184402},
    year    = {2018},
    doi     = {10.1103/PhysRevB.98.184402}
}

@article{berry_1984,
    author  = {Berry, Michael Victor},
    title   = {Quantal Phase Factors Accompanying Adiabatic Changes},
    journal = {Proc. R. Soc. Lond. A},
    volume  = {392},
    pages   = {45--57},
    year    = {1984},
    doi     = {10.1098/rspa.1984.0023}
}

@article{kittel_1951,
    author  = {Kittel, C.},
    title   = {Theory of Antiferromagnetic Resonance},
    journal = {Phys. Rev.},
    volume  = {82},
    pages   = {565},
    year    = {1951},
    doi     = {10.1103/PhysRev.82.565}
}

@article{keffer_1952,
    author  = {Keffer, F. and Kittel, C.},
    title   = {Theory of Antiferromagnetic Resonance},
    journal = {Phys. Rev.},
    volume  = {85},
    pages   = {329--337},
    year    = {1952},
    doi     = {10.1103/PhysRev.85.329}
}

@article{rezende_2019,
    author  = {Rezende, Sergio M. and Azevedo, Antonio and Rodr{\'i}guez-Su{\'a}rez, Roberto L.},
    title   = {Introduction to Antiferromagnetic Magnons},
    journal = {J. Appl. Phys.},
    volume  = {126},
    pages   = {151101},
    year    = {2019},
    doi     = {10.1063/1.5109132}
}

@article{baltz_2018,
    author  = {Baltz, V. and Manchon, A. and Tsoi, M. and Moriyama, T. and Ono, T. and Tserkovnyak, Y.},
    title   = {Antiferromagnetic Spintronics},
    journal = {Rev. Mod. Phys.},
    volume  = {90},
    pages   = {015005},
    year    = {2018},
    doi     = {10.1103/RevModPhys.90.015005}
}

@article{cham_2022,
    author  = {Cham, Thow Min Jerald and Karimeddiny, Saba and Dismukes, Avalon H. and Roy, Xavier and Ralph, Daniel C. and Luo, Yunqiu Kelly},
    title   = {Anisotropic Gigahertz Antiferromagnetic Resonances of the Easy-Axis van der Waals Antiferromagnet {CrSBr}},
    journal = {Nano Letters},
    volume  = {22},
    pages   = {6716--6723},
    year    = {2022},
    doi     = {10.1021/acs.nanolett.2c02124}
}

@article{wegrowe_2012,
    author  = {Wegrowe, J.-E. and Ciornei, M.-C.},
    title   = {Magnetization Dynamics, Gyromagnetic Relation, and Inertial Effects},
    journal = {Am. J. Phys.},
    volume  = {80},
    pages   = {607--611},
    year    = {2012},
    doi     = {10.1119/1.4709188}
}

@article{mondal_2017,
    author  = {Mondal, Ritwik and Berritta, Marco and Nandy, Ashis K. and Oppeneer, Peter M.},
    title   = {Relativistic Theory of Magnetic Inertia in Ultrafast Spin Dynamics},
    journal = {Phys. Rev. B},
    volume  = {96},
    pages   = {024425},
    year    = {2017},
    doi     = {10.1103/PhysRevB.96.024425}
}

@article{mondal_2018,
    author  = {Mondal, Ritwik and Berritta, Marco and Oppeneer, Peter M.},
    title   = {Generalisation of Gilbert Damping and Magnetic Inertia Parameter as a Series of Higher-Order Relativistic Terms},
    journal = {J. Phys.: Condens. Matter},
    volume  = {30},
    pages   = {265801},
    year    = {2018},
    doi     = {10.1088/1361-648X/aac5a2}
}

@article{quarenta_2024,
    author  = {Quarenta, Mario Gaspar and Tharmalingam, Mithuss and Ludwig, Tim and Yuan, H. Y. and Karwacki, Lukasz and Verstraten, Robin C. and Duine, Rembert A.},
    title   = {Bath-Induced Spin Inertia},
    journal = {Phys. Rev. Lett.},
    volume  = {133},
    pages   = {136701},
    year    = {2024},
    doi     = {10.1103/PhysRevLett.133.136701}
}

@article{johnsen_2025,
    author  = {Johnsen, Christian Svingen and Sudb{\o}, Asle},
    title   = {Dynamically Generated Spin Interactions and Nutational Spin Inertia in Normal Metal--Ferromagnet Heterostructures},
    journal = {Phys. Rev. B},
    volume  = {111},
    pages   = {144423},
    year    = {2025},
    doi     = {10.1103/PhysRevB.111.144423}
}

@article{kim_2023,
    author  = {Kim, Se Kwon and Tchernyshyov, Oleg},
    title   = {Mechanics of a Ferromagnetic Domain Wall},
    journal = {J. Phys.: Condens. Matter},
    volume  = {35},
    pages   = {134002},
    year    = {2023},
    doi     = {10.1088/1361-648X/acb5d8}
}
\end{document}